\documentclass[10pt,conference]{IEEEtran}
\IEEEoverridecommandlockouts

\usepackage{cite}
\usepackage{amsmath,amssymb,amsfonts}
\usepackage{algorithm}
\usepackage{algpseudocode}
\usepackage{textcomp}
\usepackage{xcolor}
\usepackage{enumitem}
\usepackage{listings}
\usepackage{float}
\usepackage{graphicx}
\usepackage{enumitem}
\usepackage{adjustbox}
\usepackage{tcolorbox}
\usepackage{svg}
\usepackage{balance}
\usepackage{booktabs}
\usepackage{url}

\newcommand{\tool}[0]{\textsc{arat-rl}}
\newcommand{\smalltt}[1]{{\footnotesize\texttt{#1}}}

\newlength{\textfloatsepsave}
\colorlet{punctcolor}{red!60!black}
\colorlet{desccolor}{green!60!black}
\colorlet{altcolor}{blue!60!black}
\colorlet{kwcolor}{teal!60!black}
\definecolor{keywordcolor}{rgb}{0.13, 0.29, 0.53}

\lstdefinelanguage{json}{
    basicstyle=\scriptsize\ttfamily\color{altcolor},
    numbers=none,
    showstringspaces=false,
    backgroundcolor=\color{gray!5},
    framexleftmargin=4pt,
    escapeinside={@}{@},
    tabsize=3,
    showstringspaces=false,
    breaklines=true,
    frame=tb,
    rulecolor=\color{black},
    literate=
    {:}{{{\color{punctcolor}{:}}}}{1}
    {,}{{{\color{punctcolor}{,}}}}{1}
    {-}{{{\color{punctcolor}{-}}}}{1}
    {[}{{{\color{punctcolor}{\textbf{[}}}}}{1}
    {]}{{{\color{punctcolor}{\textbf{]}}}}}{1},
}

\begin{document}

\title{Adaptive REST API Testing with \\Reinforcement Learning}

\author{\IEEEauthorblockN{Myeongsoo Kim}
\IEEEauthorblockA{
Georgia Institute of Technology\\
Atlanta, Georgia, USA \\
mkim754@gatech.edu}
\and
\IEEEauthorblockN{Saurabh Sinha}
\IEEEauthorblockA{
IBM Research\\
Yorktown Heights, New York, USA \\
sinhas@us.ibm.com}
\and
\IEEEauthorblockN{Alessandro Orso}
\IEEEauthorblockA{
Georgia Institute of Technology\\
Atlanta, Georgia, USA \\
orso@cc.gatech.edu}
}

\maketitle

\begin{abstract}
Modern web services increasingly rely on REST APIs. Effectively testing these APIs is challenging due to the vast search space to be explored, which involves selecting API operations for sequence creation, choosing parameters for each operation from a potentially large set of parameters, and sampling values from the virtually infinite parameter input space. Current testing tools lack efficient exploration mechanisms, treating all operations and parameters equally (i.e., not considering their importance or complexity) and lacking prioritization strategies. Furthermore, these tools struggle when response schemas are absent in the specification or exhibit variants. To address these limitations, we present an adaptive REST API testing technique that incorporates reinforcement learning to prioritize operations and parameters during exploration. Our approach dynamically analyzes request and response data to inform dependent parameters and adopts a sampling-based strategy for efficient processing of dynamic API feedback. We evaluated our technique on ten RESTful services, comparing it against state-of-the-art REST testing tools with respect to code coverage achieved, requests generated, operations covered, and service failures triggered. Additionally, we performed an ablation study on prioritization, dynamic feedback analysis, and sampling to assess their individual effects. Our findings demonstrate that our approach outperforms existing REST API testing tools in terms of effectiveness, efficiency, and fault-finding ability.
\end{abstract}


\section{Introduction}

The increasing adoption of modern web services has led to a growing reliance on REpresentational State Transfer (REST) APIs for communication and data exchange~\cite{richardson2013restful, patni2017pro}. REST APIs adhere to a set of architectural principles that enable scalable, flexible, and efficient interactions between various software components through the use of standard HTTP methods and a stateless client-server model~\cite{fielding2000architectural}. 
To facilitate their discovery and use by clients, REST APIs are often documented using various specification languages~\cite{openapi, swagger, raml, apiblueprint} that let developers describe the APIs in a structured format and provide essential information, such as the available endpoints, input parameters and their schemas, response schemas, and so on.
Platforms such as APIs Guru~\cite{apis_guru} and Rapid~\cite{rapidapi} host thousands of RESTful API documents, emphasizing the significance of formal API specifications in industry.

Standardized documentation formats, such as the OpenAPI specification~\cite{openapi}, not only facilitate the development of REST APIs and their use by clients, but also provide a foundation for the development of automated testing techniques for such APIs, and numerous such techniques and tools have emerged in recent years (e.g., ~\cite{arcuri2019restful, Corradini2022, atlidakis2019restler, karlsson2020quickrest, martin2021restest, karlsson2020automatic, zac2022schemathesis, wu2022combinatorial}). In spite of this, testing REST APIs continues to be a challenging task, with high code coverage remaining an elusive goal for automated tools~\cite{kim2022automated}. 

Testing REST APIs can be challenging because of the large search space to be explored, due to the large number of operations, potential execution orders, inter-parameter dependencies, and associated input parameter value constraints~\cite{kim2022automated, martin2019catalogue}. Current techniques often struggle when exploring this space due to lack of effective exploration strategies for operations and their parameters. In fact, existing testing tools tend to treat all operations and parameters equally, disregarding their relative importance or complexity, which leads to suboptimal testing strategies and insufficient coverage of crucial operation and parameter combinations. Moreover, these tools rely on discovering producer-consumer relationships between response schemas and request parameters; this approach works well when the parameter and response schemas are described in detail in the specification, but falls short in the common case in which the schemas are incomplete or imprecise.

To address these limitations of the state of the art, we present adaptive REST API testing with reinforcement learning (\tool), an advanced black-box testing approach. Our technique incorporates several innovative features, such as leveraging reinforcement learning to prioritize operations and parameters for exploration, dynamically constructing key-value pairs from both response and request data, analyzing these pairs to inform dependent operations and parameters, and utilizing a sampling-based strategy for efficient processing of dynamic API feedback. The primary objectives of our approach are to increase code coverage and improve fault-detection capability.

The core novelty in \tool\ is an adaptive testing strategy driven by a 
reinforcement-learning-based prioritization algorithm for exploring the space of operations and parameters. The algorithm initially determines the importance of an operation based on the parameters it uses and the frequency of their use across operations. This targeted exploration enables efficient coverage of critical operations and parameters, thereby improving code coverage. The technique employs reinforcement learning to adjust the priority weights associated with operations and parameters based on feedback, by decreasing importance for successful responses and increasing it for failed responses. The technique also assigns weights to parameter-value mappings based on various sources of input values (e.g., random, specified values, response values, request values, and default values). The adaptive nature of \tool\ gives precedence to yet-to-be-explored or previously error-prone operations, paired with suitable value mappings, which improves the overall efficiency of API exploration.

Another innovative feature of \tool\ is dynamic construction of key-value pairs. In contrast to existing approaches that rely heavily on resource schemas provided in the specification, our technique dynamically constructs key-value pairs by analyzing POST operations (i.e., the HTTP methods that create resources) and examining both response and request data. For instance, suppose that an operation takes book title and price as request parameters and, as response, produces a success status code along with a string message (e.g., ``Successfully created''). Our technique leverages this information to create key-value pairs for book title and price, upon receiving a successful response, even if such data is not present in the response. In other words, the technique takes into account the input parameters used in the request, as they correspond to the created resource. Moreover, if the service returns incomplete resources (i.e., only a subset of the data items for a given type of resource), our technique still creates key-value pairs for the information available. This dynamic approach enables \tool\ to identify resources from the API responses and requests, as well as discover hidden dependencies that are not evident from the specification alone.

Finally, \tool\ employs a simple yet effective sampling-based approach that allows it to process dynamic API feedback efficiently and adapt its exploration based on the gathered information. By randomly sampling key-value pairs from responses, our technique reduces the overhead of processing every response for each pair, resulting in more efficient testing. 

To evaluate \tool, we conducted a set of empirical studies using 10 RESTful services and compared its performance against three state-of-the-art REST API testing tools: RESTler~\cite{atlidakis2019restler}, EvoMaster~\cite{arcuri2019restful}, and Morest~\cite{liu2022morest}. We assessed the effectiveness of \tool\ in terms of coverage achieved and service failures triggered, and its efficiency in terms of valid and fault-inducing requests generated and operations covered within a given time budget. Our results show that \tool\ outperforms the competing tools for all the metrics considered---it achieved the highest method, branch, and line coverage rates, along with better fault-detection ability. Specifically, \tool\ covered 119\%, 60\%, and 52\% more branches, lines, and methods than RESTler; 37\%, 21\%, and 14\% more branches, lines, and methods than EvoMaster; and 24\%, 12\%, and 10\% more branches, lines, and methods than Morest. \tool\ also uncovered 9.3x, 2.5x, and 2.4x more bugs than RESTler, EvoMaster, and Morest, respectively. In terms of efficiency, \tool\ generated 52\%, 41\%, and 1,222\% more valid and fault-inducing requests and covered 15\%, 24\%, and 283\% more operations than Morest, EvoMaster, and RESTler, respectively.
We also conducted an ablation study to assess the individual effects of prioritization, dynamic feedback analysis, and sampling on the overall effectiveness of \tool. Our results indicate that reinforcement-learning-based prioritization contributes the most to \tool's effectiveness, followed by dynamic feedback analysis and sampling.

\begin{figure}[t]
  \centering
  \begin{minipage}{\linewidth}
    \begin{lstlisting}[
        language=json,
        basicstyle=\tiny,
        belowskip=-0.7\baselineskip
        ]
/products/{productName}/configurations/{configurationName}/features/{featureName}:
 @\color{keywordcolor}post@:
    @\color{keywordcolor}operationId@: addFeatureToConfiguration
    @\color{keywordcolor}produces@:
      - application/json
    @\color{keywordcolor}parameters@:
      - @\color{keywordcolor}name@: productName
        @\color{keywordcolor}in@: path
        @\color{keywordcolor}required@: true
        @\color{keywordcolor}type@: string
      - @\color{keywordcolor}name@: configurationName
        @\color{keywordcolor}in@: path
        @\color{keywordcolor}required@: true
        @\color{keywordcolor}type@: string
      - @\color{keywordcolor}name@: featureName
        @\color{keywordcolor}in@: path
        @\color{keywordcolor}required@: true
        @\color{keywordcolor}type@: string
    @\color{keywordcolor}responses@:
      @\color{keywordcolor}default@:
        @\color{keywordcolor}description@: successful operation
/products/{productName}/configurations/{configurationName}/features:
  @\color{keywordcolor}get@:
    @\color{keywordcolor}operationId@: getConfigurationActivedFeatures
    @\color{keywordcolor}produces@:
      - application/json
    @\color{keywordcolor}parameters@:
      - @\color{keywordcolor}name@: productName
        @\color{keywordcolor}in@: path
        @\color{keywordcolor}required@: true
        @\color{keywordcolor}type@: string
      - @\color{keywordcolor}name@: configurationName
        @\color{keywordcolor}in@: path
        @\color{keywordcolor}required@: true
        @\color{keywordcolor}type@: string
    @\color{keywordcolor}responses@:
      @\color{keywordcolor}'200'@:
        @\color{keywordcolor}description@: successful operation
        @\color{keywordcolor}schema@:
          @\color{keywordcolor}type@: array
          @\color{keywordcolor}items@:
            @\color{keywordcolor}type@: string
    \end{lstlisting}
  \end{minipage}
  \vspace*{2pt}
  \caption{A Part of Features Service's OpenAPI Specification.}
  \label{fig:oas_example}
\end{figure}

The main contributions of this work are:
\begin{itemize}
\item A novel approach for adaptive REST API testing that incorporates (1) reinforcement learning to prioritize exploration of operations and parameters, (2) dynamic analysis of request and response data to identify dependent parameters, and (3) a sampling strategy for efficient processing of dynamic API feedback.
\item Empirical results demonstrating that \tool\ outperforms state-of-the-art REST API testing tools by generating more valid and fault-inducing requests, covering more operations, achieving higher code coverage, and triggering more service failures.
\item An artifact~\cite{artifact} containing the tool, the benchmark services, and the empirical results.
\end{itemize}

The rest of this paper is organized as follows. Section~\ref{background} presents background information and a motivating example to illustrate challenges in REST API testing. Section~\ref{our_approach} describes our approach. Section~\ref{evaluation} presents our empirical evaluation and results.
Section~\ref{related_work} discusses related work and, finally, Section~\ref{conclusion} presents our conclusions and potential directions for future work.

\section{Background and Motivating Example}
\label{background}

We provide a brief introduction to REST APIs,  OpenAPI specifications, and reinforcement learning; then, we illustrate the novel features of our approach using a running example.

\subsection{REST APIs}

REST APIs are web APIs that adhere to the RESTful architectural style~\cite{fielding2000architectural}.
REST APIs facilitate communication between clients and servers by exchanging data through standardized protocols, such as HTTP~\cite{rodriguez2008restful}.
Key principles of REST include statelessness, cacheability, and a uniform interface, which simplify client-server interactions and promote loose coupling~\cite{tilkov2007brief}.
Clients communicate with web services by sending HTTP requests. These requests access and/or manipulate resources managed by the service, where a resource represents data that a client may want to create, delete, update, or access. Requests are sent to an API endpoint, identified by a resource path and an HTTP method specifying the action to be performed on the resource. The most commonly used methods are POST, GET, PUT, and DELETE, for creating, reading, updating, and deleting a resource, respectively. The combination of an endpoint and an HTTP method is called an operation. Besides specifying an operation, a request can also optionally include HTTP headers containing metadata and a body with the request's payload. Upon receiving and processing a request, the web service returns a response containing headers, possibly a body, and an HTTP status code---a three-digit number indicating the request's outcome. Specifically, 2xx status codes denote successful responses, 4xx codes indicate client errors, and status code 500 suggests server error.

\subsection{OpenAPI Specification}

The OpenAPI Specification (OAS)~\cite{openapi} is a widely adopted API description format for RESTful APIs, providing a standardized and human-readable way to describe the structure, functionality, and expected behavior of an API.
Figure~\ref{fig:oas_example} illustrates an example OAS file describing a part of the Features Service API. This example shows two API operations. The first operation, a POST request, is designed to add a feature name to a product's configuration. It requires three parameters: product name, configuration name, and feature name, all of which are specified in the path. Upon successful execution, the API responds with a JSON object, signaling that the feature has been added to the configuration. The second operation, a GET request, retrieves the active features of a product's configuration. Similar to the first operation, it requires the product name and configuration name as path parameters. The API responds with a 200 status code and an array of strings representing the active features in the specified configuration.

\subsection{Reinforcement Learning and Q-Learning}
\label{sec:backgorund-rlql}

Reinforcement learning (RL) is a type of machine learning where an agent learns to make decisions by interacting with an environment~\cite{sutton2018reinforcement}. The agent selects actions in various situations (states), observes the consequences (rewards), and learns to choose the best actions to maximize the cumulative reward over time. The learning process in RL is trial-and-error based, meaning the agent discovers the best actions by trying out different options and refining its strategy based on the observed rewards.
The agent must also decide between exploring new actions to gather more knowledge or exploiting known actions that offer the best reward based on its current understanding. The balance between exploration and exploitation is often governed by parameters, such as $\epsilon$ in the $\epsilon$-greedy strategy~\cite{sutton2018reinforcement}.

Q-learning is a widely used model-free reinforcement learning algorithm that estimates the optimal action-value function, $Q(s,a)$~\cite{watkins1992q}. The Q-function represents the expected cumulative reward the agent can obtain by taking action $a$ in state $s$ and then following the optimal policy. Q-learning uses a table to store Q-values and updates them iteratively based on the agent's experiences. 
In the learning process, the agent takes actions, receives rewards, and updates the Q-values using the Q-learning update rule, derived from the Bellman equation~\cite{sutton2018reinforcement}:

\vskip -15pt
\begin{equation}
\label{eq:qval-update}
Q(s,a) \leftarrow Q(s,a) + \alpha [r + \gamma \max_{a'} Q(s',a') - Q(s,a)]
\end{equation}
\vskip -5pt

where $\alpha$ is the learning rate, $\gamma$ is the discount factor, $s'$ is the new state after taking action $a$, and $r$ is the immediate reward received. The agent updates the Q-values to converge to their optimal values, which represent the expected long-term reward of taking each action in each state.

\subsection{Motivating Example}
\label{motivating_example}

Next, we illustrate the salient features of \tool\ using the Feature-Service specification (Figure~\ref{fig:oas_example}) as an example. 

\vskip 5pt
\noindent
\textbf{RL-based adaptive exploration}:
For the example in Figure~\ref{fig:oas_example}, to perform the operation \smalltt{addFeatureToConfiguration},
we must first create a product using a separate operation and establish a configuration for it using another operation. The sequence of operations should, therefore, be: create product, create configuration, and create feature name for the product with the specified configuration name. This example emphasizes the importance of determining the operation sequence. Our technique initially assigns weights to operations and parameters based on their usage frequency in the specification.
In this case, \smalltt{productName} is the most frequently used parameter across all operations; therefore, our technique assigns higher weights to operations involving \smalltt{productName}. Specifically, the operation for creating a product gets the highest priority.

Moreover, once an operation is executed, its priority must be adjusted so that it is not explored repeatedly, creating new product instances unnecessarily.
After processing a prioritized operation, our technique employs RL to adjust the weights in response to the API response received. If a successful response is obtained, negative rewards are assigned to the processed parameters, as our objective is to explore other uncovered operations. This method naturally leads to the selection of the next priority operation and parameter, facilitating efficient adjustments to the call sequence.

Inter-parameter dependencies~\cite{martin2019catalogue} can increase the complexity of the testing process, as some parameters might have mutual exclusivity or other constraints associated with them (e.g., only one of the parameters can be specified). RL-based exploration guided by feedback received can also help with dealing with this complexity.

\vskip 5pt
\noindent
\textbf{Dynamic construction of key-value pairs}: Existing REST API testing strategies (e.g.,~\cite{Corradini2022, atlidakis2019restler, liu2022morest}) emphasize the importance of identifying producer-consumer relationships between response schemas and request parameters. However, current tools face limitations when operations produce unstructured output (e.g., plain text) or have incomplete response schemas in their specifications. For instance, the \smalltt{addFeatureToConfiguration} operation lacks structured response data (e.g., JSON format). Despite this, our approach processes and generates key-value data \smalltt{\{productName: <value>, configurationName: <value>, featureName: <value>\}} from the request data, as the POST HTTP method indicates that a resource is created using the provided inputs.

By analyzing and storing key-value pairs identified from request and response data, 
our dynamic key-value pair construction method proves especially beneficial in cases of responses with plain-text descriptions or incomplete response schemas.
The technique can effectively uncover hidden dependencies not evident from the specification.

\vskip 5pt
\noindent
\textbf{Sampling for efficient dynamic key-value pair construction}:
API response data can sometimes be quite large
and processing every response for each key-value pair can be computationally expensive.
To address this issue, we have incorporated a samplingd strategy into our dynamic key-value pair construction method. This strategy efficiently processes the dynamic API feedback and adapts its exploration based on the gathered information while minimizing the overhead of processing every response.

\begin{figure}[t]
\centering
\includegraphics[width=\columnwidth]{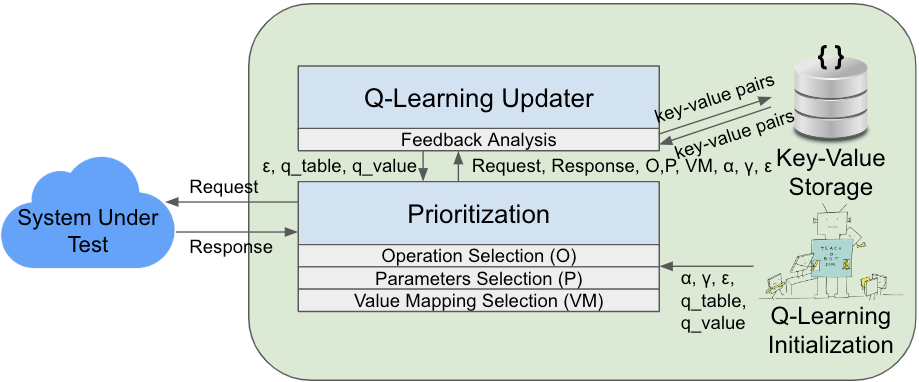}
\caption{Overview of our approach.}
\label{fig:overview}
\end{figure}

\section{Our Approach}\label{our_approach}

In this section, we introduce our Q-Learning-based REST API testing approach, which intelligently prioritizes and selects operations, parameters, and value-mapping sources while dynamically constructing key-value pairs. Figure~\ref{fig:overview} provides a high-level overview of our approach. Initially, the Q-Learning Initialization module sets up the necessary variables and tables for the Q-learning process. Q-Learning Updater subsequently receives these variables and tables and passes them to the Prioritization module, which is responsible for selecting operations, parameters, and value-mapping sources.

\tool\ then sends a request to the System Under Test (SUT) and receives a response. It also supplies the request, response, selected operation, parameters, mapped value source, and the Q-Learning parameters ($\alpha$, $\gamma$, and $\epsilon$) to Q-Learning Updater.  The feedback is analyzed with the request and response, storing key-value pairs extracted from them for future use. The Updater component then adjusts the Q-values based on the outcomes, enabling the approach to adapt its decision-making process over time. \tool\ iterates through this procedure until the specified time limit is reached. In the rest of this section, we present the details of the algorithm.

\subsection{Q-Learning Table Initialization}
The Q-Learning Table Initialization component, shown in Algorithm~\ref{alg:initialization}, is responsible for setting up the initial Q-table and Q-value data structures that guide the decision-making process throughout a testing session. Crucially, this process happens without making any API calls.

The algorithm begins by setting the learning rate $\alpha$ to~0.1, the discount factor $\gamma$ to 0.99, and the exploration rate $\epsilon$~to~0.1 (lines 2--4). These parameters control the learning and exploration process of the Q-Learning algorithm; the chosen values are the ones that are commonly recommended and used (e.g.,~\cite{qlearningex1, qlearningex2, masadeh2018reinforcement}). The algorithm then initializes the Q-table and the Q-value with empty dictionaries (lines 5--6).

\begin{algorithm}[t]
\caption{Q-Learning Table Initialization}
\scriptsize
\label{alg:initialization}
\begin{algorithmic}[1]
\Procedure{InitializeQLearning}{operations}
    \State Set learning rate ($\alpha$) to 0.1
    \State Set discount factor ($\gamma$) to 0.99
    \State Set exploration rate ($\epsilon$) to 0.1
    \State Initialize empty dictionary q\_table
    \State Initialize empty dictionary q\_value
    \For{operation in operations}
        \State operation\_id $\gets$ operation['operationId']
        \State q\_value[\text{operation\_id}] $\gets$ new dictionary
        \For{source in [S1, S2, S3, S4, S5]}
            \State q\_value[\text{operation\_id}][\text{source}] $\gets$ 0
        \EndFor
        \For{parameter in operation['parameters']}
            \State param\_name $\gets$ parameter['name']
            \If{key in q\_table}
                \State q\_table[\text{param\_name}] = q\_table[\text{param\_name}] + 1
            \Else
                \State q\_table[\text{param\_name}] = 1
            \EndIf
        \EndFor
        \For{response\_data in operation\_data.get('responses')}
            \For{key in response\_data.keys()}
                \If{key in q\_table}
                    \State q\_table[\text{key}] = q\_table[\text{key}] + 1
                \EndIf
            \EndFor
        \EndFor
    \EndFor
    \State \Return $\alpha$, $\gamma$, $\epsilon$, q\_table, q\_value
\EndProcedure
\end{algorithmic}
\end{algorithm}

The algorithm iterates through each operation in the API (lines 7--24). For each operation, it extracts the operation's unique identifier (operation\_id) and creates a new entry in the Q-value dictionary for the operation (lines 8--9). Next, it initializes the Q-value for each value-mapping source (S1--S5) to zero (lines 10--12).

The algorithm proceeds to iterate through each parameter in the operation (lines 13--20). It extracts the parameter's name (param\_name) and, if param\_name already exists in the Q-table, increments the corresponding entry by one; otherwise, it initializes the entry to one. This step builds the Q-table with counts of occurrences of each parameter.

Next, the algorithm iterates through the response data of each operation (lines 21--27). It extracts keys from the response and checks, for each key, whether it is present in the Q-table for that operation (line 18). If a key is present, the algorithm increments the corresponding entry in the Q-table by one (lines 23--25). This step populates the Q-table with the frequency of occurrence of each response key.

Finally, the algorithm returns the learning rate $\alpha$, the discount factor $\gamma$, the exploration rate $\epsilon$, the Q-table, and the Q-value (line 29). This initial setup provides the Q-Learning algorithm with basic information about the API operations and their relationships, which is further refined during testing.

\begin{algorithm}[t]
\caption{Q-Learning-based Prioritization}
\scriptsize
\label{alg:q_prioritization}
\begin{algorithmic}[1]
\Procedure{SelectOperation}{operations, q\_table}
    \State Initialize \text{max\_avg\_q\_value} $\gets -\infty$
    \State Initialize \text{best\_operation} $\gets \text{None}$
    \For{operation in operations}
        \State operation\_id $\gets$ operation['operationId']
        \State Initialize \text{sum\_q\_value} $\gets$ 0
        \State Initialize \text{num\_params} $\gets$ \text{len(operation['parameters'])}
        \For{parameter in operation['parameters']}
            \State param\_name $\gets$ parameter['name']
            \State \text{sum\_q\_value} $\gets$ \text{sum\_q\_value} + q\_table[\text{param\_name}]
        \EndFor
        \State \text{avg\_q\_value} $\gets$ \text{sum\_q\_value} / \text{num\_params}
        \If{\text{avg\_q\_value} $>$ \text{max\_avg\_q\_value}}
            \State \text{max\_avg\_q\_value} $\gets$ \text{avg\_q\_value}
            \State \text{best\_operation} $\gets$ \text{operation}
        \EndIf
    \EndFor
    \State \Return \text{best\_operation}
\EndProcedure
\end{algorithmic}
\begin{algorithmic}[1]
\Procedure{SelectParameters}{operation, $\epsilon$}
    \State Set n randomly (0 $\leq$ n $\leq$ \text{length of operation['parameters']})
    \State Initialize empty list selected\_parameters
    \If{\text{random.random()} $> \epsilon$}
        \State Sort operation['parameters'] by Q-values in descending order
        \For{i $\gets$ 0 to n $-$ 1}
            \State Append operation['parameters'][i] to selected\_parameters
        \EndFor
    \Else
        \For{param in \text{random.sample}(operation['parameters'], n)}
            \State Append param to selected\_parameters
        \EndFor
    \EndIf
    \State \Return selected\_parameters
\EndProcedure
\end{algorithmic}
\begin{algorithmic}[1]
\Procedure{SelectValueMappingSource}{operation, $\epsilon$}
    \begin{description}
        \item[Source1:] Example values in specification
        \item[Source2:] Random value generated based on parameter's type and format
        \item[Source3:] Dynamically constructed key-value pairs from request
        \item[Source4:] Dynamically constructed key-value pairs from response
        \item[Source5:] Default values (string: string, number: 1.1, integer: 1, array: [], object: \{\})
    \end{description}
    \State operation\_id $\gets$ operation['operationId']
    \State sources $\gets$ [S1, S2, S3, S4, S5]
    \If{\text{random.random()} $> \epsilon$}
    \State max\_q\_value $\gets -\infty$
    \State max\_q\_index $\gets -1$
    \For{\text{s} in \text{sources}}
        \If{\text{q\_value}[\text{operation\_id}][\text{s}] $> \text{max\_q\_value}$}
            \State max\_q\_value $\gets \text{q\_value}[\text{operation\_id}][\text{s}]$
            \State max\_q\_index $\gets \text{s}$
        \EndIf
    \EndFor
        \State \Return max\_q\_index
    \Else
        \State \Return random.randint(1, 5)
    \EndIf

\EndProcedure
\end{algorithmic}
\end{algorithm}

\begin{algorithm}[t]
\caption{Q-Learning-based API Testing}
\scriptsize
\label{alg:api_testing_algorithm}
\begin{algorithmic}[1]
\Procedure{QLearningUpdater}{response, q\_table, q\_value, selected\_op, selected\_params, $\alpha$, $\gamma$}
\State operation\_id $\gets$ selected\_op['operation\_id']
\If{response.status\_code is 2xx (successful)}
\State Extract key-value pairs from request and response
\State reward $\gets -1$
\State Update q\_value negatively
\ElsIf{response.status\_code is 4xx or 500 (unsuccessful)}
\State reward $\gets 1$
\State Update q\_value positively
\EndIf
\For{each param in selected\_params}
\For{each param\_name, param\_value in param.items()}
\State old\_q\_value $\gets$ q\_table[operation\_id][param\_name]
\State max\_q\_value\_next\_state $\gets$ max(q\_table[operation\_id].values())
\State q\_table[operation\_id][param\_name] $\gets$ old\_q\_value + $\alpha$ * (reward + $\gamma$ * (max\_q\_value\_next\_state - old\_q\_value))
\EndFor
\EndFor
\State \Return q\_table, q\_value
\EndProcedure
\end{algorithmic}
\begin{algorithmic}[1]
\Procedure{Main}{API specification}
\State Initialize $\epsilon_\mathit{max} \gets 1$
\State Initialize $\epsilon_\mathit{adapt} \gets 1.1$
\State Initialize time\_limit $\gets$ desired time limit in seconds
\State operations $\gets$ Load API specification
\State $\alpha$, $\gamma$, $\epsilon$, q\_table, q\_value $\gets$ \textsc{InitializeQLearning}(operations)
\While{time\_limit not reached}
\State operation $\gets$ \textsc{SelectOperation}(operations, q\_table) \State parameters $\gets$ \textsc{SelectParameters}(operation, $\epsilon$)
\State source $\gets$ \textsc{SelectValueMappingSource}(operation, $\epsilon$)
\State response $\gets$ Execute API request with operation, parameters, and source
\State q\_table, q\_value $\gets$ \textsc{QLearningUpdater}(response, q\_table, q\_value, operation, parameters, $\alpha$, $\gamma$)
\State $\epsilon \gets \min(\epsilon_\mathit{max}, \epsilon_\mathit{adapt} * \epsilon)$
\EndWhile
\EndProcedure
\end{algorithmic}
\end{algorithm}

\subsection{Q-Learning-based Prioritization}
In this step, \tool\ prioritizes API operations and selects the best parameters and value-mapping sources based on their Q-values. We present Algorithm~\ref{alg:q_prioritization} (\textit{SelectOperation}, \textit{SelectParameters}, and \textit{SelectValueMappingSource}) to describe the prioritization process.

The \textit{SelectOperation} procedure is responsible for selecting the best API operation to exercise next. The algorithm initializes variables to store the maximum average Q-value and the best operation (lines 2--3). It then iterates through each operation, calculating the average Q-value for the operation's parameters (lines 4--17). The operation with the highest average Q-value is selected as the best operation (line 15).

The \textit{SelectParameters} procedure selects a subset of parameters for the chosen API operation. This selection is guided by the exploration rate $\epsilon$. If a random value is greater than $\epsilon$, the algorithm selects the top $n$ parameters sorted by their Q-values in descending order (lines 4--8); otherwise, it randomly selects $n$ parameters from the operation's parameters (lines 9--12). Finally, the selected parameters are returned (line 14).

The \textit{SelectValueMappingSource} procedure is responsible for selecting the value-mapping source for the chosen API operation. The technique leverages five sources of values.

\begin{itemize}

\item Source 1 (example values in specification): These values are provided in the API documentation as examples for a parameter. We consider three types of OpenAPI keywords that can specify example values: \smalltt{enum}, \smalltt{example}, and \smalltt{description}~\cite{openapi}. The OpenAPI documentation~\cite{openapi} states that users can specify example values in the \textit{description} field, and a recent study also shows the importance of leveraging example values from descriptions~\cite{kim2023enhancing}. However, such examples are not provided in a structured format but as natural-language text. To extract example values from the \textit{description} field, we create a list containing each word
in the text, as well as each quoted phrase.

\item Source 2 (random value generated based on parameter's type, format, and constraints): This source generates random values for each parameter based on its type, format, and constraints. To generate random values, we utilize Python's built-in \textit{random} library~\cite{random}. For date and date-time formats, we employ the \textit{datetime} library~\cite{datetime} to randomly select dates and times. If the parameter has a regular expression pattern specified in the API documentation, we generate the value randomly using the \textit{rstr} library~\cite{rstr}. When a minimum or maximum constraint is present, we pass it to the \textit{random} library to ensure that the generated values adhere to the specified constraints. This approach allows \tool\ to explore a broader range of values compared to the example values provided in the API specification.

\item Source 3 (dynamically constructed key-value pairs from request): This source extracts key-value pairs from dynamically constructed request key-value pairs. We employ Gestalt pattern matching~\cite{difflib} (also known as Ratcliff/Obershelp pattern recognition~\cite{paul}) to identify the key most similar to the parameter name. Gestalt matching is a lightweight but effective technique that calculates the similarity between two strings by identifying the longest common subsequences and recursively comparing the remaining unmatched substrings. This technique aids in discovering producer-consumer relationships.

\item Source 4 (dynamically constructed key-value pairs from response): Similar to Source 3, this source obtains key-value pairs from dynamically constructed response key-value pairs. We use the same Gestalt pattern matching approach~\cite{difflib} to identify the key, further assisting in the identification of producer-consumer relationships.

\item Source 5 (default values): This source uses predefined default values for each data type: ``string'' for strings, 1.1~for numbers, 1~for integers: 1, [] for arrays, and \{\} for objects. These default values can be useful for testing how the API behaves when it receives the simplest or common forms of inputs; such default values are used by other tools as well (e.g.,~\cite{atlidakis2019restler}).

\end{itemize}

Similar to the \textit{SelectParameters} procedure, the selection of the value-mapping source is guided by $\epsilon$. If a random value is greater than $\epsilon$, the algorithm selects the mapping source with the highest Q-value for the chosen operation (lines~4--6). This helps the algorithm focus on the most-promising mapping sources based on prior experience. Otherwise, the algorithm randomly selects a mapping source from the available sources (line~8). This randomness ensures that the algorithm occasionally explores less-promising mapping sources to avoid getting stuck in a suboptimal solution.

\subsection{Q-Learning-based API Testing}
In this step, \tool\ executes the selected API operations with the selected parameters and value-mapping sources, and updates the Q-values based on the response status codes. Algorithm~\ref{alg:api_testing_algorithm} (\textit{QLearningUpdater} and \textit{Main}) describes the API testing process and the update of Q-values using the learning rate ($\alpha$) and discount factor ($\gamma$).

The \textit{QLearningUpdater} procedure updates the Q-values based on the response status codes. It first extracts the operation ID from the selected operation (line~2). If the response status code indicates success (2xx), the algorithm extracts key-value pairs from the request and response, assigns a reward of~$-1$, and updates the Q-values negatively (lines~3--6). If the response status code indicates an unsuccessful request (4xx or 500), the algorithm assigns a reward of~1 and updates the Q-values positively (lines~7--10). The Q-values are updated for each parameter in the selected parameters using the Q-learning update rule (Equation~\ref{eq:qval-update} in \S\ref{sec:backgorund-rlql} (lines~11--16), and the updated Q-values are returned (line~18).

The \textit{Main} procedure orchestrates the Q-Learning-based API testing process. It initializes the exploration rate ($\epsilon$), its maximum value ($\epsilon_\mathit{max}$), its adaptation factor ($\epsilon_\mathit{adapt}$), and the time budget for testing (lines~2--4). The API specification is loaded and the Q-Learning table is initialized (lines~5--6). The algorithm then enters a loop that continues until the time limit is reached (line~7). In each iteration, the best operation, selected parameters, and selected mapping source are determined (lines~8--10). The API operation is executed with the selected parameters and mapping source, and the response is obtained (line~11). The Q-values are then updated based on the response (line~12), and the exploration rate ($\epsilon$) is updated (line~13).

By continuously updating the Q-values based on the response status codes and adapting the exploration rate, Q-Learning-based API testing process aims to effectively explore the space of API operations and parameters. 

\section{Evaluation}\label{evaluation}

\begin{figure*}[htbp]
    \centering
    \includegraphics[width=\textwidth]{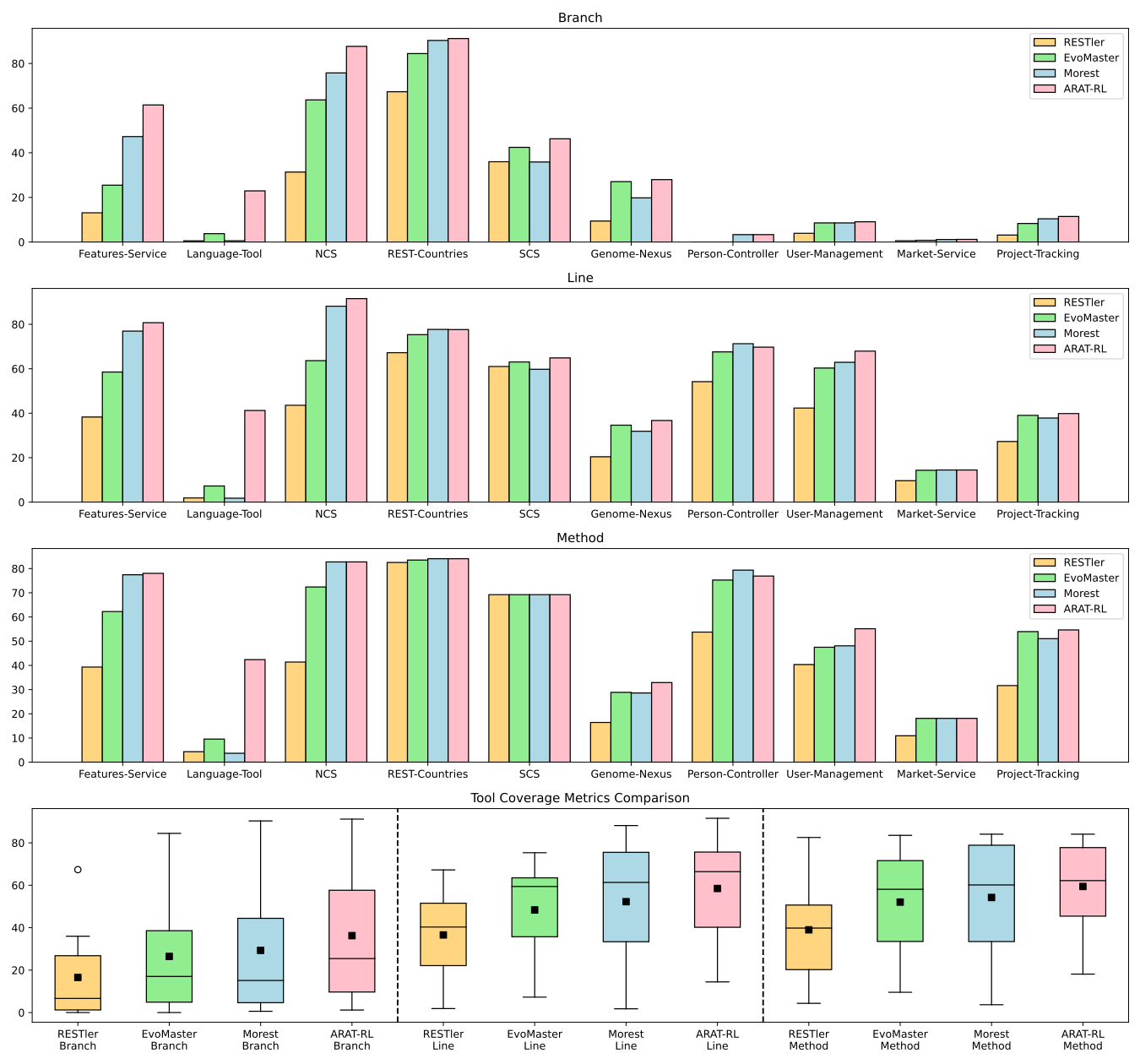}
    \vspace*{-22pt}
    \caption{Branch, line, and method coverage achieved by the tools on the benchmark services.}
    \vspace*{-14pt}
    \label{fig:effectiveness_comparison}
\end{figure*}

In this section, we present the results of empirical studies conducted to assess \tool. Our evaluation aims to address the following research questions:

\begin{enumerate}
\item \textbf{RQ1}: How does \tool\ compare with state-of-the-art REST API testing tools for in terms of code coverage?
\item \textbf{RQ2}: How does the efficiency of \tool, measured in terms of valid and fault-inducing requests generated and operations covered within a given time budget, compare to that of other REST API testing tools?
\item \textbf{RQ3}: In terms of error detection, how does \tool\ perform in identifying 500 responses in REST APIs compared to state-of-the-art REST API testing tools?
\item \textbf{RQ4}: How do the main components of \tool---prioritization, dynamic key-value pair construction, and sampling---contribute to its overall performance?
\end{enumerate}

\subsection{Experiment Setup}

We performed our experiments using Google Cloud E2 machines, each equipped with 24-core CPU and 96 GB of RAM. We created a machine image containing all the services and tools in our benchmark. For each experiment, we deleted and recreated the machines using this image to minimize potential dependency issues. Each machine hosted all the services and tools under test, but we ran one tool at a time during the experiments. We monitored CPU and memory usage throughout the testing process to ensure that the testing tools were not affected by a lack of memory or CPU resources.

To evaluate the effectiveness and efficiency of our approach, we compared its against three state-of-the-art tools: EvoMaster~\cite{arcuri2019restful}, RESTler~\cite{atlidakis2019restler}, and Morest~\cite{liu2022morest}. We selected 10 RESTful services from a recent REST API testing study~\cite{kim2022automated} as our benchmark. We explain the selection process of these tools and services next.

\subsubsection*{Testing Tools Selection}

As a preliminary note, because \tool\ is a black-box approach, we considered black-box tools only in our evaluation. We believe that adding white-box tools to the evaluation would result in an unfair comparison, as these tools leverage information about the code, rather than just information in the specification, to generate test inputs. 

We identified an initial set of 10 tools based on a recent study~\cite{kim2022automated}. 
From this list, we chose (the black-box version of) EvoMaster~\cite{arcuri2019restful} and RESTler~\cite{atlidakis2019restler}. 
EvoMaster employs an evolutionary algorithm for automated test case generation and was the best-performing tool in that study and in another recent comparison~\cite{zhang2022open}. Its strong performance makes it an appropriate candidate for comparison. RESTler adopts a grammar-based fuzzing approach to explore APIs. It is a well-established tool in the field and, in fact, the most popular REST API testing tool in terms of GitHub stars.

Recently, two new tools have been published. Morest~\cite{liu2022morest} has been shown to have superior results compared to EvoMaster. We, therefore, included Morest in our set of tools for comparison, as it could potentially outperform the other tools. The other recent tool, RestCT, was also considered for inclusion in our evaluation. However, we encountered failures while running it. We contacted the authors, who confirmed the issues and said they will work on resolving them.

\subsubsection*{RESTful Services Selection}

As benchmarks for our evaluation, we selected 10 out of 20 RESTful services from a recent study~\cite{kim2022automated}. We had to exclude 10 services for various reasons.
Specifically, we omitted the News service, developed by the author of one of our baseline tools (EvoMaster), to avoid possible bias. 
Problem Controller and Spring Batch REST were excluded because they require specific domain knowledge to generate meaningful tests, so using them provides limited information about the tools.
We excluded Erc20 Rest Service and Spring Boot Actuator because some APIs in these services did not provide valid responses due to external dependencies being updated without corresponding updates in the service code.
Proxyprint, OCVN, and Scout API could not be included due to authentication issues that prevented them from generating meaningful responses.
Finally, we excluded CatWatch and CWA Verification because of their restrictive rate limits, which slowed down the testing process and made it impossible to collect results in a reasonable amount of time.

Our final set of services consisted of Features Service, LanguageTool, NCS, REST Countries, SCS, Genome Nexus, Person Controller, User Management Microservice, Market Service, and Project Tracking System.

\begin{table*}[t]
\centering
\caption{Comparison of operations covered and valid and failure-inducing requests generated (2xx and 500 status codes) by the tools.}
\vspace*{-7pt}
\label{tab:efficiency}
\begin{adjustbox}{width=\textwidth}
\begin{tabular}{l|c|rrr|r|rrr|c|rrr|c|rrr}
\toprule
& \multicolumn{4}{c|}{{ARAT-RL}} & \multicolumn{4}{c|}{{Morest}} & \multicolumn{4}{c|}{{EvoMaster}} & \multicolumn{4}{c}{{RESTler}} \\
\cmidrule{2-17}
& \#operations & \multicolumn{3}{c|}{{\#requests}} & \#operations & \multicolumn{3}{c|}{{\#requests}} & \#operations & \multicolumn{3}{c|}{{\#requests}} & \#operations & \multicolumn{3}{c}{{\#requests}} \\
Service & covered & 2xx+500 & 2xx & 500 & covered & 2xx+500 & 2xx & 500 & covered & 2xx+500 & 2xx & 500 & covered & 2xx+500 & 2xx & 500 \\ \midrule
Features Service & 18 & 95,479 & 43,460 & 52,019 & 18 & 103,475 & 4,920 & 98,555 & 18 & 113,136 & 33,271 & 79,865 & 17 & 4,671 & 1,820 & 2,851 \\
Language Tool & 2 & 77,221 & 67,681 & 9,540 & 1 & 1,273 & 1,273 & 0 & 2 & 22,006 & 17,838 & 4,168 & 1 & 32,796 & 32,796 & 0 \\
NCS & 6 & 62,618 & 62,618 & 0 & 5 & 18,389 & 18,389 & 0 & 2 & 61,282 & 61,282 & 0 & 2 & 140 & 140 & 0 \\
REST Countries & 22 & 36,297 & 35,486 & 811 & 22 & 8,431 & 7,810 & 621 & 16 & 9,842 & 9,658 & 184 & 6 & 259 & 255 & 4 \\
SCS & 11 & 115,328 & 115,328 & 0 & 11 & 110,147 & 110,147 & 0 & 10 & 66,313 & 66,313 & 0 & 10 & 5,858 & 5,857 & 1 \\
Genome Nexus & 23 & 15,819 & 14,010 & 1,809 & 23 & 32,598 & 10,661 & 21,937 & 19 & 8,374 & 8,374 & 0 & 1 & 182 & 182 & 0 \\
Person Controller & 12 & 101,083 & 47,737 & 53,346 & 11 & 104,226 & 10,036 & 94,190 & 12 & 91,316 & 37,544 & 53,772 & 1 & 167 & 102 & 65 \\
User Management Microservice & 21 & 44,121 & 13,356 & 30,765 & 17 & 1,111 & 948 & 163 & 18 & 29,064 & 13,003 & 16,061 & 4 & 79 & 64 & 15 \\
Market Service & 12 & 29,393 & 6,295 & 23,098 & 6 & 1,399 & 394 & 1,005 & 5 & 10,697 & 4,302 & 6,395 & 2 & 1,278 & 0 & 1,278 \\
Project Tracking System & 53 & 23,958 & 21,858 & 2,100 & 42 & 14,906 & 12,904 & 2,002 & 43 & 15,073 & 13,470 & 1,603 & 3 & 72 & 65 & 7 \\ \midrule
Average & 18 & 60,132 & 42,783 & 17,349 & 15.6 & 39,595 & 17,748 & 21,847 & 14.5 & 42,710 & 26,505 & 16,205 & 4.7 & 4,550 & 4,128 & 422 \\
\bottomrule
\end{tabular}
\end{adjustbox}
\end{table*}

\subsubsection*{Result Collection}

We ran each testing tool with the time budget of one hour per execution, as a previous study~\cite{kim2022automated} showed that code coverage achieved by these tools tends to plateau within this duration. To accommodate randomness, we replicated the experiments ten times and calculated the average metrics across the runs.

Data collection for code coverage and status codes was done using JaCoCo~\cite{jacoco} and Mitmproxy~\cite{mitmproxy}, respectively. We focused on identifying unique instances of 500 codes, which indicate server-side faults. The methodology was as follows:

\begin{enumerate}
\item \textbf{Stack Trace Collection:} For services that provided stack traces with 500 errors, we collected these traces, treating each unique trace as a separate fault. In the majority of cases, the errors we collected fall into this category.

\item \textbf{Response Text Analysis:} In the absence of stack traces, we analyzed the response text. After removing unrelated components (e.g., timestamps), we classified unique instances of response text linked to 500 status codes as individual faults.
\end{enumerate}

This systematic approach allowed us to compile a comprehensive and unique tally of faults for our analysis.

\subsection{RQ1: Effectiveness}

To answer RQ1, we compared the tools in terms of branch, line, and method coverage achieved. Figure~\ref{fig:effectiveness_comparison} presents the results of the study. The bar charts represent the coverage achieved by each tool for each RESTful service, whereas the boxplot at the bottom summarizes of each tool's performance on the three coverage metrics across all the services.

As the boxplot illustrates, \tool\ consistently outperforms the other tools in all three coverage metrics over the subject services.
Looking at the performance breakdown by services (shown in the bar charts), \tool\ performed the best (or was tied as the best-performing tool) for all services in terms of branch coverage (with one ties), eight of the 10 services in terms of line coverage (with no tie), and nine of the 10 services in terms of method coverage (with four ties). In cases where \tool\ did not achieve the best results, it still achieved similar coverage rates as the best-performing tool.


\tool\ is especially effective when there is operation dependency, parameter dependency, and value-mapping dependency. For example, the highest coverage gains occurred for Language Tool, which has a complex set of inter-parameter and value-mapping dependencies. Meanwhile, \tool\ struggles with semantic parameters. For instance, its effectiveness was the lowest on Market Service, although it still matched the best-performing tool on this service. The reason for this is that the service requires input data, such as address, email, name, password, and phone number in specific formats, but \tool\ failed to generate valid values for these. Consequently, it was unable to create market users and then use that information for other operations in producer-consumer relationships.

On average, \tool\ attained 36.25\% branch coverage, 58.47\% line coverage, and 59.42\% method coverage. In comparison, Morest, which exhibited the second-best performance, reached an average of 29.31\% branch coverage, 52.27\% line coverage, and 54.24\% method coverage. Thus, the coverage gain of \tool\ over Morest is 23.69\% for branch coverage, 11.87\% for line coverage, and 9.55\% for method coverage. EvoMaster and RESTler yield lower average coverage rates on all three metrics, with respective results of 26.45\%, 48.37\%, and 52.07\% for EvoMaster and 16.54\%, 36.58\%, and 38.99\% for RESTler for branch, line, and method coverage. The coverage gains of \tool\ compared to EvoMaster is 37.03\% for branch coverage, 20.87\% for line coverage, and 14.13\% for method coverage; compared to RESTler, the gains are 119.17\% for branch coverage, 59.83\% for line coverage, and 52.42\% for method coverage.

These results provide evidence that our technique can more effectively explore REST services, achieving superior code coverage, compared to existing tools, and demonstrate its potential in addressing the challenges in REST API testing.

\begin{tcolorbox}[left=3pt, right=3pt, top=2pt, bottom=2pt]
\tool\ consistently outperforms RESTler, EvoMaster, and Morest in terms of branch, line, and method coverage across the subject services.
However, \tool\ can struggle with parameters that require inputs in specific formats. 
\end{tcolorbox}

\begin{table}[t]
    \centering
    \caption{Total faults detected by the tools over 10 runs.}
    \vspace*{-5pt}
    \label{tab:faults}
    \resizebox{0.9\columnwidth}{!}{
    \begin{tabular}{l|cccc}
    \toprule
        Service & {RESTler} & {EvoMaster} & {Morest} & {ARAT-RL} \\ \midrule
        {Features Service}    & \phantom010 & \phantom010 & \phantom010 & \phantom010 \\
        {Language Tool}       & \phantom0\phantom00 & \phantom048 & \phantom0\phantom00 & 122 \\
        {NCS}                 & \phantom0\phantom00 & \phantom0\phantom00 & \phantom0\phantom00 & \phantom0\phantom00 \\
        {REST Countries}      & \phantom0\phantom09 & \phantom010 & \phantom010 & \phantom010 \\
        {SCS}                 & \phantom0\phantom03 & \phantom0\phantom00 & \phantom0\phantom00 & \phantom0\phantom00 \\
        {Genome Nexus}        & \phantom0\phantom00 & \phantom0\phantom00 & \phantom0\phantom05 & \phantom010 \\
        {Person Controller}   & \phantom058 & 221 & 274 & 943 \\
        {User Management Microservice}     & \phantom010 & \phantom010 & \phantom0\phantom08 & \phantom010 \\
        {Market Service}      & \phantom010 & \phantom010 & \phantom010 & \phantom010 \\
        {Project Tracking System}    & \phantom010 & \phantom010 & \phantom010 & \phantom010 \\ \midrule
        {Total}                 & 110 & 319 & 327 & 1125 \\
        \bottomrule
    \end{tabular}
    }
\end{table}

\subsection{RQ2: Efficiency}

To address RQ2, we compared \tool\ to Morest, EvoMaster, and RESTler in terms of the number of (1) valid and fault-inducing requests generated (as indicated by HTTP status codes 2xx and 500, respectively) and (2) operations covered within a given time budget. Although efficiency is not only dependent on these metrics, due to factors such as API response time, we feel that they still represent meaningful proxies because they indicate the extent to which the tools are exploring the API and identifying faults.

Table~\ref{tab:efficiency} shows these metrics for the 10 subject services. For each service, the table lists the number of operations covered and the number of requests made under the categories 2xx+500 (sum of 2xx code and 500 status code), 2xx, and 500, for each of the four tools. In the last row, the table presents the average number of operations covered and requests made by each tool across all the services.

In the testing time budget of one hour, \tool\ generated more valid and failure-inducing requests, resulting in more exploration of the testing search space. Specifically, \tool\ generated 60,132 valid and fault-inducing requests on average, which is 52.01\% more than Morest (39,595 requests), 40.79\% more than EvoMaster (42,710 requests), and 1222\% more than RESTler (4,550 requests). 

This difference in the number of requests can be attributed to \tool’s approach of processing only a sample of key-value pairs from the response, rather than the entire response. By focusing on sampling key-value pairs, \tool\ efficiently identifies potential areas of improvement, contributing to a more effective REST API testing process.

Moreover, \tool\ covered more operations on average (18 operations) compared to Morest (15.6 operations), EvoMaster (14.5 operations), and RESTler (4.7 operations). This indicates that \tool\ efficiently generates more requests in a given time budget, which leads to covering more API operations, contributing to more comprehensive testing process.

\begin{tcolorbox}[left=3pt, right=3pt, top=2pt, bottom=2pt]
Given a one-hour testing time budget, \tool\ generates 52.01\%, 40.79\%, and 1222\% more valid and fault-inducing requests than Morest, EvoMaster, and RESTler, respectively. Additionally, it covers 15.38\%, 24.14\%, and 282.98\% more operations than these tools.
\end{tcolorbox}

\subsection{RQ3: Fault-Detection Capability}

Table~\ref{tab:faults} presents the total number of faults detected by each tool across 10 runs for the RESTful services in our benchmark. We note that this cumulative count might include multiple detection of the same fault in different runs. 
For clarity in discussion, we provide the average faults detected per run in the text below.
As indicated, \tool\ exhibits the best fault-detection capability, uncovering on average 113 faults over the services. In comparison, Morest and EvoMaster detected 33 and 32 faults on average, respectively, whereas RESTler found the least---11 faults on average.
\tool\ thus uncovered 9.3x, 2.5x, and 2.4x more faults than RESTler, EvoMaster, and Morest, respectively.

Comparing this data against the data on 500 response codes from Table~\ref{tab:efficiency}, we can see that, although \tool\ generated 20.59\% fewer 500 responses, it found 250\% more faults than Morest. Compared to EvoMaster, \tool\ generated 7.06\% more 500 responses, but also detected 240\% more faults. These results suggest that irrespective of whether \tool\ generates more or fewer error responses, it is better at uncovering more unique faults. 

\tool's superior fault-detection capability is particularly evident in the Language Tool and Person Controller services, where it detected on average 12 and 94 faults, respectively. What makes this significant is that these services have larger sets of parameters. Our strategy performed well in efficiently prioritizing and testing various combinations of these parameters to reveal faults. For example, Language Tool's main operation \smalltt{/check}, which checks text grammar, has 11 parameters. Similarly, Person Controller's main operation \smalltt{/api/person}, which creates/modifies person instances, has eight parameters. In contrast, the other services' operations usually have three or fewer parameters.

\tool\ intelligently tries various parameter combinations with a reward system in Q-learning because it gives negative rewards for the parameters in the successful requests. This ability to explore various parameter combinations is a significant factor in revealing more bugs, especially in services with a larger number of parameters. These results indicate that \tool's reinforcement-learning-based approach can effectively discover faults in REST services.

\begin{tcolorbox}[left=3pt, right=3pt, top=2pt, bottom=2pt]
\tool\ exhibits superior fault-detection ability, uncovering 9.3x, 2.5x, and 2.4x more bugs than RESTler, EvoMaster, and Morest, respectively. This is mainly attributed to its intelligent RL-based exploration of various parameter combinations in services with large number of parameters.
\end{tcolorbox}

\begin{table}[t]
    \centering
    \caption{Results of the ablation Study.}
    \vspace*{-7pt}
    \begin{adjustbox}{width=\columnwidth}
    \label{tab:ablation_study}
    \setlength{\tabcolsep}{4pt} 
    \begin{tabular}{l|cccc}
        \toprule
        & {Branch} & {Line} & {Method} & {Faults Detected} \\ \midrule
        {ARAT-RL}                      & 36.25 & 58.47 & 59.42 & 112.10 \\
        {ARAT-RL (no prioritization)} & 28.70 (+26.3\%) & 53.27 (+9.8\%) & 55.51 \phantom0(+7\%) & 100.10 (+12\%) \\
        {ARAT-RL (no feedback)}       & 32.69 (+10.9\%) & 54.80 (+6.9\%) & 56.09 (+5.9\%) & 110.80 (+1.2\%) \\
        {ARAT-RL (no sampling)}       & 34.10 \phantom0(+6.3\%) & 56.39 (+3.7\%) & 57.20 (+3.9\%) & 112.50 (-0.4\%) \\
        \bottomrule
    \end{tabular}
    \end{adjustbox}
\end{table}

\subsection{RQ4: Ablation Study}

To address RQ4, we conducted an ablation study to assess the impact of the main novel components of  \tool\ on its performance. We compared the performance of \tool\ to three variants: (1) \tool\ without prioritization, (2) \tool\ without dynamic key-value construction from feedback, and (3) \tool\ without sampling. Table~\ref{tab:ablation_study} presents the results of this study.

As illustrated in the table, the removal of any component results in reductions in branch, line, and method coverage, as well as the number of faults detected. Eliminating the prioritization component leads to the most substantial decline in performance, with branch, line, and method coverage decreasing to 28.70\%, 53.27\%, and 55.51\%, respectively, and the number of detected faults dropping to 100. This highlights the critical role of the prioritization mechanism in \tool's effectiveness.

The absence of feedback and sampling components also negatively affects performance. Without feedback, \tool's branch, line, and method coverage decreases to 32.69\%, 54.80\%, and 56.09\%, respectively, and the number of found faults is reduced to 110.80. Likewise, without sampling, branch, line, and method coverage drops to 34.10\%, 56.39\%, and 57.20\%, respectively, while the number of found faults experiences a slight increase to 112.50, which may be attributed to random variations. 

These findings emphasize that each component of \tool\ is essential for its overall effectiveness. The prioritization mechanism, feedback loop, and sampling strategy work together to optimize the tool's performance in terms of code coverage and fault detection capabilities.

\begin{tcolorbox}[left=3pt, right=3pt, top=2pt, bottom=2pt]
Each component of \tool\---prioritization, feedback, and sampling---contributes to \tool's overall effectiveness. The prioritization mechanism, in particular, plays a significant role in enhancing \tool's performance in code coverage achieved and faults detected.
\end{tcolorbox}

\subsection{Threats to Validity}

In this section, we address potential threats to our study's validity and the steps taken to mitigate them. Internal validity is influenced by tool implementations and configurations.
To minimize these threats, we used the latest tool versions and used default options. Some testing tools might have randomness, but we addressed this issue by running each tool 10 times and computing the average results.

Our data-collection process relies on an automated script. Despite thorough testing, there remains the possibility that the script could contain errors, potentially affecting our results. To mitigate this risk, we conducted several rounds of pilot testing on a representative subset of our services before applying the script to the full study.

External validity is affected by the selection of RESTful services, the limited number of RESTful services tested, impacting the generalizability of our findings. We tried to ensure a fair evaluation by selecting a diverse set of 10 services, but we will try to have a larger and more diverse set in future work.

Construct validity concerns the metrics and tool comparisons used. Metrics such as branch, line, and method coverage, number of requests, and 500 status codes as faults, although commonly, may not capture the test tools' full quality. Additional metrics, such as mutation scores, could provide a better understanding of tool effectiveness. We measured efficiency by considering the number of meaningful requests generated by the tools, including valid and fault-inducing ones, as well as the number of operations each tool covered within a given time budget. While these metrics give us some perspective on a tool's performance, they do not consider all possible factors. Other aspects, such as the response times from the services, may also significantly impact overall efficiency.

Including more tools, services, and metrics in future studies would allow a more comprehensive evaluation of our technique's performance.

\section{Related Work}\label{related_work}

In this section, we provide an overview of related work in automated REST API testing, requirements based test case generation, and reinforcement learning based test case generation.

\textbf{Automated REST API testing}: EvoMaster~\cite{arcuri2019restful} is a technique that offers both white-box and black-box testing modes. It uses evolutionary algorithms for test case generation, refining tests based on its fitness function and checking for 500 status code. Other black-box techniques include various tools with different strategies. RESTler~\cite{atlidakis2019restler} generates stateful test cases by inferring producer-consumer dependencies and targets internal server failures. RestTestGen~\cite{Corradini2022} exploits data dependencies and uses oracles for status codes and schema compliance. QuickREST~\cite{karlsson2020quickrest} is a property-based technique with stateful testing, checking non-500 status codes and schema compliance. Schemathesis~\cite{zac2022schemathesis} is a tool designed for detecting faults using five types of oracles to check response compliance in OpenAPI or GraphQL web APIs via property-based testing. RESTest~\cite{martin2021restest} accounts for inter-parameter dependencies, producing nominal and faulty tests with five types of oracles. Morest~\cite{liu2022morest} uses a dynamically updating RESTful-service Property Graph for meaningful test case generation. RestCT~\cite{wu2022combinatorial} employs Combinatorial Testing for RESTful API testing, generating test cases based on Swagger specifications.

Open-source black-box tools such as Dredd~\cite{dredd}, fuzz-lightyear~\cite{fuzz-lightyear}, and Tcases~\cite{tcases} offer various testing capabilities. Dredd~\cite{dredd} tests REST APIs by comparing responses with expected results, validating status codes, headers, and body content. Using Swagger for stateful fuzzing, fuzz-lightyear~\cite{fuzz-lightyear} detects vulnerabilities in micro-service ecosystems, offers customized test sequences, and alerts on unexpected successful attacks. Tcases~\cite{tcases} is a model-based tool that constructs an input space model from OpenAPI specifications, generating test cases covering valid input dimensions and checking response status codes for validation.

\textbf{Requirements based test case generation}: In requirements based testing, notable works include ucsCNL~\cite{barros2011ucscnl}, which uses controlled natural language for use case specifications, and UML Collaboration Diagrams~\cite{badri2004use}. Requirements by Contracts~\cite{nebut2003requirements} proposed a custom language for functional requirements, while SCENT-Method~\cite{ryser1999scenario} employed a scenario-based approach with statecharts. SDL-based Test Generation~\cite{tahat2001requirement} transformed SDL requirements into EFSMs, and RTCM~\cite{yue2015rtcm} introduced a natural language-based framework with templates, rules, and keywords. These approaches typically deal with the goal of generating test cases as accurately as possible according to the intended requirements. In contrast, \tool\ focuses on more formally specified REST APIs, which provide a structured basis for testing and effectively explore and adapt during the testing process to find faults.

\textbf{Reinforcement learning based test case generation}: Several recent studies have investigated the use of reinforcement learning in software testing, focusing primarily on web applications and mobile apps. These challenges often arise from hidden states, whereas in our approach, we have access to all states through the API specification but face more constraints on operations, parameters, and mapping values. Zheng et al.\cite{zheng2021webrl} proposed an automatic web client testing approach utilizing curiosity-driven reinforcement learning. Pan et al.\cite{pan2020androidrl} introduced a similar curiosity-driven approach for testing Android applications. Koroglu et al.\cite{koroglu2018qbe} presented QBE, a Q-learning-based exploration method for Android apps. Mariani et al.\cite{mariani2021autoblacktest} proposed AutoBlackTest, an automatic black-box testing approach for interactive applications. Adamo et al.\cite{adamo2018androidrl} developed a reinforcement learning-based technique specifically for Android GUI testing. Vuong and Takada\cite{vuong2018androidrl} also applied reinforcement learning to automated testing of Android apps. Köroğlu and Sen~\cite{koroglu2020androidrl} presented a method for generating functional tests from UI test scenarios using reinforcement learning for Android applications.

\section{Conclusion and Future Work}\label{conclusion}

We introduced \tool, a reinforcement-learning-based approach for the automated testing of REST APIs. To assess the effectiveness, efficiency, and fault-detection capability of our approach, we compared its performance to that of three state-of-the-art testing tools. Our results show that \tool outperformed all three tools considered in terms of branch, line, and method coverage achieved, requests generated, and faults detected. We also conducted an ablation study, which highlighted the important role played by the novel features of our technique---prioritization, dynamic key-value construction based on response data, and sampling from response data to speed up key-value construction---in enhancing \tool's performance.

In future work, we will further improve \tool's performance by addressing the limitations we observed in recognizing semantic constraints and generating inputs with specific formatting requirements (e.g., email addresses, physical addresses, or phone numbers). 
We plan to investigate ways to combine the RESTful-service Property Graph (RPG) approach of Morest with our feedback model to better handle complex operation dependencies.
We will also study the impact of RL prioritization over time. 
Additionally, we plan to develop an advanced technique based on multi-agent reinforcement learning. This approach will rely on multiple agents that apply natural language analysis on the text contained in server responses, utilize a large language model to sequence operations more efficiently, and employ an enhanced value generation technique using extensive parameter datasets from Rapid API~\cite{rapidapi} and APIs-guru~\cite{apis_guru}. Lastly, we aim to detect a wider variety of bugs, including bugs that result in status codes other than 500.


\section*{Acknowledgments}

We thank our anonymous reviewers for their invaluable feedback and the developers of the testing tools we employed in our empirical evaluation (EvoMasterBB, Morest, and RESTler) for making their tools accessible. This research was partially supported by NSF, under grant CCF-0725202, DARPA, under contract N66001-21-C-4024, DOE, under contract DE-FOA-0002460, and gifts from Facebook, Google, IBM Research, and Microsoft Research. This publication reflects the views only of the authors; the sponsoring agencies cannot be held responsible for such views and any use which may be made of the information contained in the paper.

\section*{Data availability}

The artifact associated with this submission, which includes code, datasets, and other relevant materials, is available in our GitHub repository~\cite{artifact}. The artifact has been designed to reproduce the experiments presented in the paper and support further research in this domain. While we have archived the submission version of our artifact on Zenodo~\cite{zenodo}, we recommend visiting our GitHub repository to access the most recent version of our tool.

\balance
\bibliographystyle{IEEEtran}
\bibliography{paper}

\end{document}